\documentstyle[twocolumn,aps,psfig]{revtex}
\begin{document}
\draft
\twocolumn[\hsize\textwidth\columnwidth\hsize\csname @twocolumnfalse\endcsname
%
%
% Title Page
%

\title{Phase Separation Induced by Orbital Degrees of Freedom \\
in Models for Manganites with Jahn-Teller Phonons}
%Phase Diagram of
%the Kondo Model for Manganites Including Jahn-Teller
%Phonons}
%: \\
%Orbital-driven 
%Orbital
%
%Driven by the 
%Orbital Degrees of Freedom}
% Diagram and Optical Conductivity}

\author{ S. Yunoki, A. Moreo, and E. Dagotto}

\address{National High Magnetic Field Lab and Department of Physics,
Florida State University, Tallahassee, FL 32306}

\date{\today}
\maketitle

\begin{abstract}

The two-orbital Kondo model with classical Jahn-Teller
phonons is studied using Monte Carlo techniques. The observed
phase diagram is rich, and includes a novel regime of phase separation
induced by the orbital degrees of freedom. Experimental consequences of
our results are discussed. In addition, the 
optical conductivity $\sigma(\omega)$ of the model
is  presented. It is shown to have several
similarities with
experimental measurements for manganites.

\end{abstract}
\vskip2pc]
\narrowtext

%
% Introduction
%
 The  properties of doped manganites are currently
under much investigation due to the dramatic
decrease in their resistivity when the spins
order ferromagnetically by
lowering the temperature $T$ or applying a magnetic 
field~\cite{jin}. This effect is caused by a metal-insulator
transition associated with the magnetic ordering.
The existence of the ferromagnetic (FM) phase is
understood based on the double-exchange (DE) mechanism~\cite{zener}.
However, experiments on manganites
have revealed a 
complicated phase diagram that also includes 
charge-ordered and antiferromagnetic (AF) phases~\cite{co}.
This rich structure is beyond the DE ideas and a more
refined approach  is 
needed to understand
these compounds.

%Recently, $layered$ manganite compounds 
%${\rm La_{1.2} Sr_{1.8} Mn_2 O_7}$
%have also been synthesized~\cite{moritomo} with properties similar to
%those of their 3D counterparts.
% Strong correlations are
%important for transition metal oxides, and, thus, theoretical guidance is
%needed to understand the behavior of manganites and for the design of new
%experiments. 

Since the 1950s the  1-orbital  FM Kondo model for manganites
has been widely studied.
However, it is only recently that
its computational analysis
started, and 
 surprises have already been observed~\cite{yunoki,yunoki2,also}. 
In particular, the transition from
the undoped spin-AF regime to
the 
spin-FM regime at finite hole-density 
occurs through phase
separation (PS),
instead of
through a canted state as  believed before. 
A growing body of experimental results indeed indicate the existence of PS
in  manganites~\cite{cox,bao}, in agreement with the theoretical 
calculations. It is conceivable that PS tendencies above the
FM critical temperature $T^{FM}_c$
could be important to explain the
colossal magnetoresistance of doped manganites.
%These  results show the importance of numerical techniques
%to obtain unbiased results in this context.
%, and thus simple models such as
%the ferromagnetic Kondo may contain important aspects of the 
%physics behind these curious compounds.

However, in spite of its rich phase diagram
 the 1-orbital Kondo model 
is incomplete for a full description of $\rm Mn$-oxides. For instance, 
dynamical Jahn-Teller (JT) distortions are also important~\cite{millis},
and a proper description of the
recently observed orbital
order~\cite{mura} obviously needs at least two orbitals.
Such a  multi-orbital model with JT phonons is  
nontrivial~\cite{millis2},
%, and
%thus far it has been studied 
%only using the dynamical mean-field approximation
and the previous experience with the 1-orbital case suggests that 
 a computational analysis is crucial to understand its properties.
In addition, it is
conceptually interesting to analyze whether 
PS~\cite{yunoki,yunoki2} exists also in multi-orbital
models.

It is precisely the purpose of this paper to report the first
computational
study of a 2-orbital model for manganites including 
JT phonons. The results show a rich phase diagram including
a novel regime of PS induced by the $orbital$, rather
than the spin, degrees of freedom (DOF). 
The Hamiltonian used here has three contributions
$H_{KJT} = H_K + H_{JT} + H_{AF}$. The first term is 
$$
H_K = -\sum_{{\bf \langle ij \rangle}\sigma a b} t_{ab}
(c^\dagger_{{\bf i} a \sigma} c_{{\bf j} b \sigma} + h.c.)
- J_H \sum_{{\bf i}a \alpha \beta} { {{\bf S}_{\bf i}}\cdot{
c^\dagger_{{\bf i}a \alpha} {\bf \sigma}_{\alpha \beta} c_{{\bf i}a \beta}     } },
\eqno{(1)}
$$
\noindent where ${\bf \langle ij \rangle}$ denotes nearest-neighbor
lattice sites,
$J_H > 0$ is the Hund coupling, 
%the hopping amplitudes $t_{ab}$
%are described below, 
$a,b=1,2$ are the two $e_g$-orbitals,
%$d_{x^2 - y^2}$ and $d_{z^2 - r^2/3}$, 
the $t_{2g}$-spins ${\bf S}_{\bf
i}$ are assumed to be classical 
(with $|{\bf S}_{\bf i}| = 1$)
since their actual value in $\rm Mn$-oxides (3/2) is large~\cite{approx},
and the rest of the notation is standard. 
None of the results
described below depends crucially on the set $\{ t_{ab} \}$
 selected~\cite{hopping}. Throughout the paper the energy units are
chosen such that $t_{11}=1$ in the $x$-direction. In addition,
since $J_H$ is large in
the real manganites, here it will be  fixed to $8$ 
%(largest
%scale in the problem) 
unless otherwise stated. Finally, the $e_g$-density
$\langle n \rangle$ is adjusted using
a chemical potential $\mu$.

The coupling with JT-phonons is through~\cite{millis}
$$
H_{JT} = \lambda \sum_{{\bf i} a b \sigma} c^{\dagger}_{{\bf i} a
\sigma}
Q^{ab}_{\bf i} c_{{\bf i} b
\sigma}
+ {{1}\over{2}} \sum_{\bf i} ( {Q^{(2)}}^2_{\bf i} + {Q^{(3)}}^2_{\bf i}),
\eqno{(2)}
$$
\noindent where $Q^{11}_{\bf i} = -Q^{22}_{\bf i} = Q^{(3)}_{\bf i}$, and
$Q^{12}_{\bf i} = Q^{21}_{\bf i} = Q^{(2)}_{\bf i}$. These phonons are assumed
to be classical. This approximation has been used and discussed in
previous literature~\cite{millis}, where it was concluded that at
temperatures of the order of the critical ones (room
temperature), or a sizable fraction
of them, the use of classical phonons captures the important physics of
the model~\cite{comm555}. 
%The reader is referred to Ref.~\cite{millis} for further
%details. 
Certainly at very low-temperatures the quantum character of phonons is
important, but this is not the range of temperatures
explored in the present paper. Note that here $T=1/10$ is about
200-300K~\cite{yunoki}. 
%
% which substantially simplifies the computational study.
%and it is a reasonable first approximation towards the determination of
%the phase diagram of Eq.(1).
Finally, a small coupling between the $t_{2g}$-spins is needed to 
account for the AF character of the real materials even
when all $\rm La$ is replaced by $\rm Ca$. This classical
Heisenberg term is
$H_{AF} = J' \sum_{\bf \langle ij \rangle} {{{\bf S}_{\bf i}}\cdot{{\bf S}_{\bf j}}},
$
where $J'$ is fixed to $0.05$
throughout the paper, a value compatible with experiments~\cite{perring}. 
To study $H_{KJT}$
a Monte Carlo (MC) technique was used. The trace over the
$e_g$-electrons is carried out exactly using library subroutines 
for a fixed background of
$t_{2g}$-spins and phonons. This background is selected based on
a Metropolis MC procedure~\cite{yunoki}.
The CPU time of the technique
grows rapidly with the number of sites $L$,
%due to the Exact Diagonalization
%needed in the fermionic sector for a fixed spin-phonon configuration,
but the method has the important advantage that it does not have
sign-problems at any $T$~\cite{yunoki}.
Finally, to analyze orbital correlations the pseudopin operator
${\bf T}_{\bf i} = {{1}\over{2}} \sum_{{\sigma}ab} 
c^{\dagger}_{{\bf i}a \sigma} {\bf \sigma}_{ab} c_{{\bf i} b \sigma}$
is used, while for spin correlations the operator is 
standard. The Fourier-transform of the pseudospin correlations is
defined as $T({\bf q}) = {{1}\over{L}} \sum_{\bf l,m} e^{i{{\bf
q}\cdot{\bf (l-m)}} } \langle 
{ {{\bf T}_{\bf m}}\cdot{{\bf T}_{\bf l}} } \rangle $,
with a similar definition for the spin structure factor
$S({\bf q})$.
% ($L=$ number of sites).

Consider first the limit $\langle n \rangle = 1.0$
(undoped manganites).
Fig.1a shows $T(q)$ and $S(q)$
at representative momenta 
%$q=0$ and $\pi$
vs. $\lambda$. For small $\lambda$
%electron-phonon coupling 
%the results are similar to those at $\lambda = 0.0$,
%namely 
a large $S(0)$
indicates a tendency to  spin-FM order induced by DE (as
in the qualitatively similar 1-orbital
problem at $\langle n \rangle = 0.5$~\cite{yunoki}).
The small values of $T(q)$ imply that in this regime the
orbitals  remain
disordered.
When the coupling reaches $\lambda_{c1} \sim 1.0$, the rapid increase of
$T(\pi)$ now suggests that the ground state has a tendency to form a
 $staggered$ (or ``antiferro'')
 orbital pattern, with the spins remaining FM aligned
since $S(0)$ is large~\cite{mermin}.
The existence of this phase was discussed before, but
using multi-orbital Hubbard models 
%with Coulomb interactions and
without phonons~\cite{orbital}. Our results show that it can
be induced by JT phonons. 
%an
%orbital-AF and spin-FM phase
%can also be 
As the coupling increases further, another transition 
at $\lambda_{c2} \sim 2.0$ occurs 
to a spin-AF orbital-FM state ($S(\pi)$ and $T(0)$ are
large). In this region a 1-orbital
approximation is suitable. Studying the spin and orbital correlations in
real-space leads to the same conclusions as discussed here.
%qualitatively correct. 

%%%%%
\begin{figure}[htbp]
\vspace{-0.7cm}
\centerline{\psfig{figure=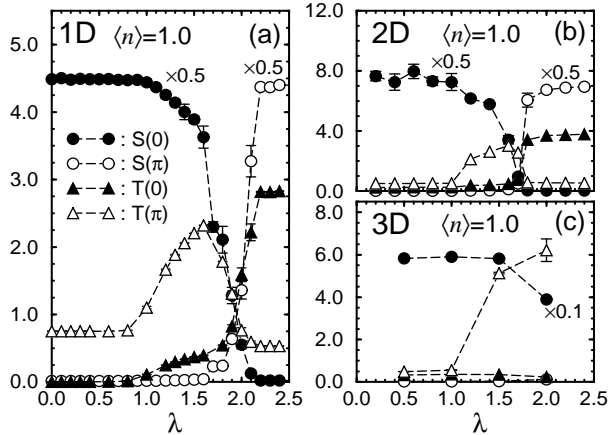,width=8.5cm,angle=-90}}
\vspace{0.2cm}
\caption{
(a) $T(q)$ and $S(q)$ vs $\lambda$, working at $\langle n \rangle =1.0$,
$T=1/75$, 
$J_H =8$, $J'=0.05$, and in 1D with $10$ sites. 
$\{ t_{ab} \}$ correspond to set $T_1$ (see [13]);
(b) Same as (a) but for a $4^2$
 cluster, $T=1/50$, and hopping $T_3$ ($T_4$) in the $y$ ($x$)
direction.
$q= 0 ( \pi )$ denotes $(0,0)$ ($(\pi,\pi)$); 
(c) Same as (a) but for a $4^3$ cluster,
$T=1/50$, the 3D 
hopping amplitudes of Ref.~[13], and $J_H = \infty$.
$q= 0 ( \pi )$ denotes $(0,0,0)$ ($(\pi,\pi,\pi)$).
}
%\vspace{0.2cm}
\end{figure}
%%%%%

The three regimes of Fig.1a can be understood in the limit 
where $\lambda$ and $J_H$ are the largest scales, and using
$t_{12}=t_{21}=0$, $t_{11}=t_{22}=t$  for simplicity. For parallel
spins with orbitals split in a staggered (uniform) pattern, the energy
per site
at lowest order in $t$ is $\sim -t^2/\Delta$ ($\sim 0$), where
$\Delta$ is the orbital splitting. For antiparallel spins with
uniform (staggered) orbital splitting, the energy is $\sim -t^2/2J_H$
($\sim -t^2/(2J_H + \Delta)$). Then, when $\Delta <  2 J_H$ (``intermediate''
$\lambda$s), a spin-FM orbital-AF
order dominates, while as $\lambda$ grows further a transition to a
spin-AF orbital-FM ground state is expected.
%Note that this reasoning is actually valid in any dimension.
%To confirm
%this prediction, 
This reasoning is dimension independent, as the results
for a 2D cluster in Fig.1b show.
%in Fig.1b results using a 2D cluster are shown. Indeed
%the qualitative behavior is very similar to the 1D results.
In 3D (Fig.1c)
%, where studies on $4^3$ clusters can only be done at
 and $J_H=\infty$
% to reduce the number of DOF, 
at least two of the regimes of Fig.1a-b have
been identified~\cite{comm66}.
It was also observed that 
the behavior in  Fig.1a-b does not change when other sets
$\{ t_{ab} \}$ are used~\cite{hopping}.
%Then, it is concluded that
%the $\langle n \rangle = 1$ phase diagram reported here
%is qualitatively dimension-independent. In addition, it was observed
%that the phases shown in Fig.1a exist for a wide variety of
%hopping matrices $t_{ab}$.
%In Fig.2a the actual real-space pseudospin correlations are shown for
%three representative values of $\lambda$. In agreement with Fig.1a,
%the weak-coupling regime has a small pseudospin correlation
%length, while the intermediate (strong) coupling
% regime has staggered (uniform) order~\cite{mermin}. 
%The behavior of the spin correlations (not shown) is standard
%i.e. when $S(0)$ ($S(\pi)$) is large they are robust and uniform (staggered).

The next issue to be explored are the transport properties
in the three regimes 
at $\langle n \rangle = 1$. The algorithm used here allows us to
calculate {\it real-time} dynamical responses accurately, including  
the optical conductivity $\sigma(\omega > 0)$,
since all the eigenvectors in
the fermionic sector for a given spin and phonon
 configuration are
obtained exactly~\cite{yunoki2}. From the sum-rule,
$e_g$ kinetic-energy,
and the integral of $\sigma(\omega > 0)$,
the $\omega = 0$ Drude-weight $D_W$ can be obtained.
In Fig.2a, $D_W$ is shown for several sizes.
%It is clear that 
$D_W$ vanishes at $\lambda_{c1}$
signaling a {\it metal-insulator} transition (MIT). 
Here the insulating phase is
spin-FM and orbital-AF, while
the metallic one is spin-FM and orbital-disordered~\cite{comm60}.
The density of
states (DOS) for $\lambda > \lambda_{c1}$ was also calculated and it
presents a clear gap at the Fermi level.
Although finite-size studies for $\rm D > 1$ are
difficult, the qualitative shape of $D_W$ vs $\lambda$ on  $4^2$ and
$4^3$ clusters was found to be the same as in 1D and, thus, it is likely
that the
MIT exists also in $D=2$ and $3$.
%all dimensions of interest.

Consider now the influence of hole doping on the $\langle n
\rangle =1.0$ phase diagram. The first issue to be addressed is the stability
of other densities as $\mu$ is varied. Fig.2b shows $\langle n
\rangle $ vs $\mu$ in
the intermediate-$\lambda$ regime. 
It is remarkable
that $two$ regions of unstable densities exist
below some critical temperature $T^{PS}_c$ (roughly 
$\sim 1/20$, see inset Fig.2b). Similar conclusions
were reached  using the Maxwell's construction~\cite{yunoki2}. 
Over $10^6$ MC sweeps
at each $\mu$
were needed for convergence near the unstable regions. 
These instabilities 
signal the existence of PS in the $H_{KJT}$ model. At 
low-density there is separation between
 an (i) empty $e_g$-electron band with AF-ordered $t_{2g}$-spins 
and a (ii) metallic
spin-FM orbital-FM phase.
%, smoothly connected
% to the small-$\lambda$ region of Fig.1a. This regime of PS is
%analogous to the spin-driven PS found at low-density in the 1-orbital 
%problem~\cite{yunoki2}.
%On the other hand,
In the unstable region near $\langle n \rangle =1.0$ 
%is $not$ contained in
%the 1-orbital problem. Here 
PS is between the phase
(ii) mentioned above,
%described in the previous paragraph, 
and (iii) the insulating 
%intermediate-$\lambda$ phase
%of Fig.1a with 
spin-FM and orbital-AF phase described in Fig.1a~\cite{maekawa}. 
The driving force
for this novel regime of PS are the orbital DOF, since
the spins are uniformly ordered in both phases involved. 
Studying $\langle n \rangle$ 
vs $\mu$, for $\lambda < \lambda_{c1}$
only PS at small densities is observed, while
for $\lambda > \lambda_{c2}$ the PS close to $\langle n \rangle = 1$
%is similar to the same phenomenon observed in the 1-orbital problem since
%it 
involves  a spin-AF orbital-FM phase~\cite{yunoki}.

%The
%existence of frequent tunneling events characteristic of unstable
%densities is shown in Fig.3a. 

%%%%%
\begin{figure}[htbp]
\vspace{-0.7cm}
\centerline{\psfig{figure=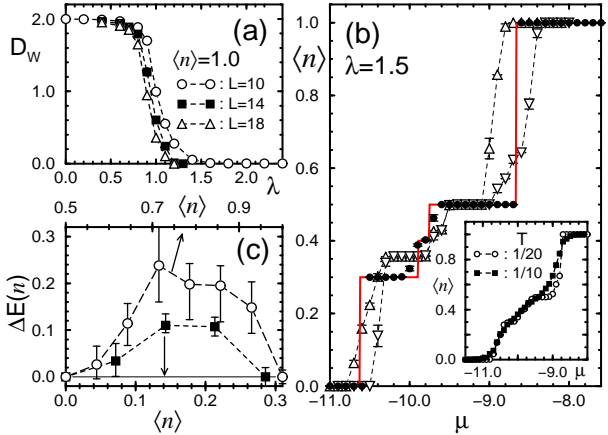,width=8.5cm,angle=-90}}
\vspace{0.2cm}
\caption{
(a) $D_W$ vs $\lambda$ for several chains ($T=1/75$);
(b) $\langle n \rangle$ vs $\mu$ at $\lambda = 1.5$, $L=10$, and
$T=1/40$ (solid circles).
%The discontinuities near $\langle n \rangle =1.0$
%and $0.0$ show the existence of unstable densities. 
The solid line is
obtained from the Maxwell's construction. The triangles are results also at
$\lambda =1.5$ and $T=1/40$, but using 14 sites and only $2 \times 10^4$ MC
sweeps to show the appearance of {\it hysteresis} loops
as in a first-order transition. The inset shows the
$T$-dependence of the results at $L=10$;
(c) $\Delta E(n)$ (see text) vs $\langle n \rangle$ 
showing a negative curvature
characteristic of PS. Results at large (open circles) 
and small (full squares) densities are shown, on a $L=14$ cluster
with periodic boundary conditions and the couplings of (b).
}
%\vspace{0.2cm}
\end{figure}
%%%%%

To confirm that the discontinuity in $\langle n \rangle$ vs $\mu$ 
corresponds to phase separation, in Fig.2c the ground state
energy is provided at several densities. The results indeed have the $negative$ curvature
characteristic of PS, both at large and small densities.
To accommodate the two important density regimes in the same plot,
the  energies in Fig.2c are defined as $\Delta E(n) = E(n) - E_0(n)$, where $E(n)$
is the actual ground state energy obtained as explained in
Ref.~\cite{yunoki2}, and $E_0(n)$ is a straight line (zero curvature)
that joins the energies of
 the two (stable) extremal densities of both the low- and high-density
regimes. Further confirmation of the PS tendencies was
obtained by inspecting the MC time-evolution of the density, as $\mu$
is varied. While at the critical $\mu$ frequent tunneling events were
observed, at other $\mu$'s the time-evolution was smooth.

Results in the limit $J_H = \infty$ using a $L=22$ site cluster have
also been obtained (see Fig.3a). Once again, a discontinuity in $\langle
n \rangle$ vs $\mu$ was found at large and small densities, 
correlated with a negative curvature in
the energy (not shown). Fig.3a helps in clarifying that the plateaus at
densities, e.g., between 0.3 and 0.5 in Fig.2c are caused by the intrinsic
discreteness of the clusters used here. As $L$ grows, this fine structure
disappears. See, e.g., the smoothly varying
density between $\langle n \rangle \sim 0.25$ and $\sim 0.55$ in Fig.3a. However,
the discontinuities at large and small $\langle n \rangle$ remain
strong, as they should if there is PS in the problem.
Note that
similar results as found in 1D appear also in studies of 2D systems (Fig.3b).

%Also note that in 2D (Fig.3b) the results are very
%similar as observed in Figs.2c and 3a.

%%%%%
\begin{figure}[htbp]
\vspace{-0.5cm}
\centerline{\psfig{figure=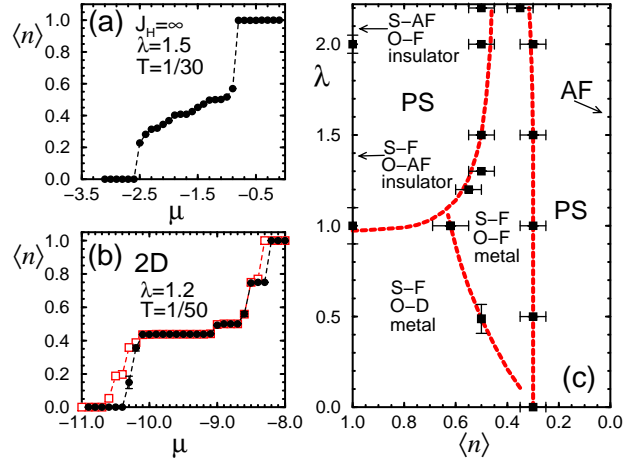,width=8.5cm,angle=-90}}
\vspace{0.2cm}
\caption{
(a) $\langle n \rangle$ vs $\mu$ in the limit $J_H = \infty$, using
$\lambda = 1.5$, $J'=0.05$, $T=1/30$, and a 22-site chain;
(b) $\langle n \rangle$ vs $\mu$ at $\lambda = 1.2$ using a $4^2$
cluster and at $T=1/50$. The two sets of points that produce the
hysteresis loops are obtained by increasing
and decreasing $\mu$ using $\sim 10^4$ sweeps at each $\mu$;
(c) Phase diagram of $H_{KJT}$ at $J_H=8.0$, $J'=0.05$,
and using set $T_1$ for  $\{ t_{ab} \}$.
$\rm S-F$ and $\rm S-AF$ denote regimes with FM- and AF-spin 
order, respectively. $\rm O-D$, $\rm O-F$, and $\rm O-AF$ represent states with
disordered, uniform and staggered orbital order, respectively.
PS means phase separation.
}
%\vspace{0.2cm}
\end{figure}
%%%%%

Then, based on the information discussed thus far, supplemented by
other MC measurements, the phase diagram of the 1D $H_{KJT}$
model is given
in Fig.3c. The metallic spin-FM region contains two regimes: one
ferro-orbital ordered and the other orbitally disordered, as deduced
from the behavior of pseudospin correlations, the mean-value of the
pseudospin operators, and the probability of double occupancy of the
same site with different orbitals. 
%The two PS regimes are shown, together with the
%metallic spin-FM region. 
The results are similar for several
$\{ t_{ab} \}$ sets~\cite{hopping}. 
Our simulations suggest that the qualitative
shape of Fig.3c should be valid also in $D=2$ and $3$.

Consider now $\sigma(\omega)$.
Experimental studies for a variety of manganites
%in $Nd_{0.7} Sr_{0.3} Mn O_3$, 
%$La_{0.7} Ca_{0.3} Mn O_3$, and $La_{0.7} Sr_{0.3} Mn O_3$ 
reported
a broad peak at $\omega \sim 1eV$ (for hole doping $x > 0.2$ and 
$T > T^{FM}_c$)~\cite{kaplan,jung}.
%as the transition from a JT-split 
%$\rm Mn^{3+}$ $e_g$-level to an unsplit $\rm Mn^{4+}$ neighboring
%site
%~\cite{oki,jung}. 
At room-$T$ 
there is negligible weight near $\omega=0$,
but as $T$ is reduced the $1eV$-peak shifts to smaller energies, gradually
transforming into a Drude response well below $T^{FM}_c$. The
finite-$\omega$
peak can be identified even inside the FM
phase. The coherent spectral weight is only a small fraction
of the total. Other features at larger energies $\sim 3eV$
involve transitions between the $J_H$-split bands and the $O$-ions.
%In addition, Jung et al.~\cite{jung}
%interpreted the $1eV$ feature at room-$T$ as composed of two
%peaks due to intra- and inter-atomic transitions in JT-distorted environments.

In Fig.4a, $\sigma(\omega)$ for the $H_{KJT}$ model is shown 
at $\langle n \rangle = 0.7$ and several temperatures 
near the unstable PS
 region of Fig.3c (weight due to $J_H$
split bands is not shown, but it appears at higher energy).
Here the FM spin correlation length 
grows rapidly with the lattice size for $T^* \leq 0.05t$, which can be
considered as the ``critical'' temperature. 
Both at high- and intermediate-$T$ a broad
peak is observed at $\omega \sim 1$, smoothly evolving to
lower energies as $T$ decreases. The peak can be identified below
$T^*$ as in 
experiments~\cite{kaplan,jung}. 
Eventually as the temperature is further reduced, $\sigma(\omega)$ is
dominated by a Drude peak.
The $T$-dependence shown in
Fig.4a is achieved at this $\lambda$ and $\langle n \rangle$ 
by a combination of a finite-$\omega$ phonon-induced broad feature
that loses weight, and a
Drude response that grows as $T$ decreases (for smaller $\lambda$s, 
the two peaks can be distinguished even at the lowest temperature shown
in Fig.4a).
The similarity with
experiments suggests that real
manganites may have couplings close to an unstable 
region in parameter space.
In the inset, $D_W$
vs $T$ is shown. Note that
%: (1) even at the lowest-$T$ of the
%simulation, $D_W$ is only 74\% of the total weight, and (2) at $T \sim T^*$,
$D_W$ vanishes suggesting a MIT, probably
 due to polaron localization. Results for the 1-orbital case
are smoother.
%, with no indications of a singularity.
%(even at $T> T^*$ the spin-spin
%correlations are robust).
%These features are in agreement with experiments, since the
%manganite ``normal''
%state is an insulator.  

%%%%%
\begin{figure}[htbp]
\vspace{-0.5cm}
\centerline{\psfig{figure=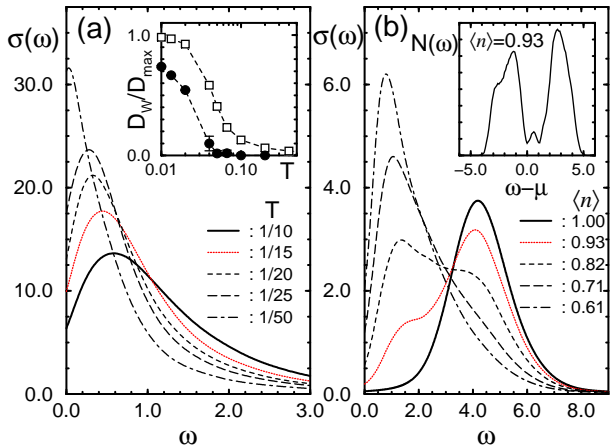,width=8.5cm,angle=-90}}
\vspace{0.2cm}
\caption{
(a) $\sigma(\omega)$
at $\lambda=1.0$, $\langle n \rangle = 0.7$, and $L=20$, and several
temperatures. 
The inset shows $D_W$ vs $T$ for both the $H_{KJT}$ (circles) and the
1-orbital model (squares) of Ref.~[4] 
(the latter at $\langle n \rangle = 0.65$). $D_W$ is
normalized to its maximum value at $T=0.01$; 
(b) $\sigma(\omega)$ vs $\omega$ parametric with
$\langle n \rangle$ at $\lambda = 1.5$, $T=1/10$, and $L=16$ (results
for $L=10$ are similar).
The inset shows the lower $J_H$-split
DOS at $\langle n \rangle = 0.93$.
%(only the lower $J_H$-split band is shown).
In (a) and (b) a
$\delta$-function broadening $\epsilon = 0.25$ was used, as well as set
$T_1$ for the hopping amplitudes.
}
%\vspace{0.2cm}
\end{figure}
%%%%%

A similar good agreement with experiments was observed working
in the regime of the orbitally-induced PS but at a temperature above
$T^{PS}_c$. Here 
 the broad feature observed at high-$T$ in Fig.4a moves to 
higher energies (Fig.4b) since $\lambda$ has increased.
At the temperature of
the plot the system is an insulator at $\langle n \rangle =
1$, but as hole carriers are added a second peak at lower energies
develop, in addition to a weak Drude peak (which carries, e.g.,
 just $1\%$ of the total weight at $\langle n \rangle = 0.61$). This 
feature at high-$T$ is reminiscent of recent experimental
 results~\cite{jung} where a two-peak 
structure was observed at room-$T$ and several
 densities. Similar results were obtained on $4^2$
clusters. In Fig.4b the peak at large-$\omega$
 is caused by
phononic effects since its position was found to grow 
rapidly with $\lambda$ ($\Delta \sim 2 \lambda \langle ( {Q^{(2)}}^2_{\bf i}
+ {Q^{(3)}}^2_{\bf i}  )^{1/2} \rangle $). 
It corresponds to intersite transitions between $\rm Mn^{3+}$ JT-split states.
The lower energy structure is compatible with a $\rm Mn^{3+}$-$\rm Mn^{4+}$
transition~\cite{warning}. The inset of Fig.4b shows the DOS
 of the system.
The two peaks above $\mu$ are responsible for the features found
in $\sigma(\omega)$. This interpretation is the same
as given in Ref.~\cite{millis2} at $D=\infty$.

Summarizing, here the first computational study of the
2-orbital Kondo model including Jahn-Teller phonons was reported. The phase diagram
includes regions of phase separation both at large and small $e_g$-density.
One of them corresponds to a novel regime where PS is driven by the
orbital, rather than spin, degrees of freedom. 
Coulomb interactions will break the large
regions involved in PS for the pure $H_{KJT}$ model into small islands of one
phase embedded into the other. X-ray diffraction 
measurements should observe a coexistence of patterns characteristics of
the two extreme stable phases, if the PS is static. If the process is
more dynamical the Bragg peaks should be broad.
Recalling that our spin-FM phase at $\langle n \rangle = 1$ and $D=1,2$ is also
compatible with A-type spin-AF order, several experimental
results~\cite{cox} are in agreement with the tendencies to PS
 discussed in this paper. In the limit of small
$\langle n \rangle$, the PS observed here could be transformed 
by long-range Coulomb interactions into a
charge-ordered state. Note that the PS observed in
electron-doped ${\rm Sr_2 Mn O_4}$~\cite{bao} 
is also compatible with our results~\cite{radaelli}.

%We thank  P. G. Radaelli and S-W. Cheong for discussions on this subject.
The authors are supported by the 
NSF grant DMR-9520776. 

%Agreement between the present calculations and the experimental measured
% $\sigma(\omega)$ was also here reported. 

%discussed in Ref....., although further work is needed to
%understand its origin. The inset of Fig.5b reveals a similar behavior on
%2D clusters.

%-------------------------

%Since
%results for the $e_g$ electrons bandwidth range
%from $BW \sim 1~eV$~\cite{bandwidth} to $BW \sim 4 eV$~\cite{sarma},
%producing a hopping $t = BW/12$ between $0.08$ and $0.33~eV$.

%------------------

\medskip

\vfil


\begin{references}
\bibitem{approx} This approximation was shown to be
accurate in Ref.~\cite{yunoki}.
\bibitem{comm555} Note that estimations of $T^{FM}_c$ using quantum and classical phonons
lead to very similar results~\cite{millis}.
\bibitem{mermin} 
$H_{KJT}$ is not invariant under arbitrary 
pseudospin-space rotations and, thus, the Mermin-Wagner theorem does not apply 
to the orbital
correlations.
\bibitem{comm66}  Studies at
larger values of $\lambda$'s are 
too costly in CPU time.
\bibitem{comm60} In Ref.~\cite{millis2}, a MIT at $\lambda \sim 1$ was
also found but it was not associated with orbital order.
\bibitem{warning} Although its dependence with $\lambda$ was weak.
\end{references}
\end{document}